# Classifying soft elastic lattices using higher-order homogenization

BASILE AUDOLY

Laboratoire de Mécanique des Solides
CNRS, Institut Polytechnique de Paris
91120 Palaiseau, France

CLAIRE LESTRINGANT

Institut Jean Le Rond d'Alembert,
Sorbonne Université, CNRS,
75005 Paris, France

HUSSEIN NASSAR

Department of Mechanical and Aerospace Engineering,
University of Missouri, Columbia, MO 65211, USA

We propose a methodology for the homogenization of periodic elastic lattices that covers the case of unstable lattices, having affine (macroscopic) or periodic (microscopic) mechanisms. The singular cell problems that are encountered when a periodic mechanism is present are naturally solved by treating the amplitude $\theta(X)$ of the mechanism as an enrichment variable. We use asymptotic second-order homogenization to derive an effective energy capturing both the strain-gradient effect $\nabla \varepsilon$ relevant to affine mechanisms, and the $\nabla \theta$ regularization relevant to periodic mechanisms, if any is present. The proposed approach is illustrated with a selection of lattices displaying a variety of effective behaviors. It follows a unified pattern that leads to a classification of these effective behaviors.

## 1. INTRODUCTION

Solving unit cell problems under periodic boundary conditions is standard procedure when it comes to computing the effective elastic moduli of an inhomogeneous solid. In doing so, the solid is typically presumed stable. In linear elasticity, stability is as crucial as it is common: indeed, any mixture of elastic phases with non-vanishing elastic moduli is stable. Notwithstanding, *the purpose of the present paper is to chart, classify and exemplify cases where stability fails and to clarify what becomes of the effective behavior and of its effective moduli*. It can be anticipated then that the solids of interest will necessarily contain voids, or structural elements with vanishing moduli. The paper limits itself to the simplest of such solids, namely to periodic pin-jointed trusses hereafter referred to as *lattices*.

In *unstable* lattices, there exist deformations called *mechanisms* that alter distances between nodes without stretching bonds [38, 30, 12, 21]. Sometimes, by cutting a finite sample free out of a lattice, a number of mechanisms are created and are confined to "loose ends" [28]. These mechanisms are inconsequential as far as the effective behavior of the lattice is concerned. More relevant to our purposes are mechanisms defined by the *Cauchy-Born* rule,

$$v_i^{\mathrm{m}} = \varepsilon^{\mathrm{m}} \cdot X_i + \xi_{b(i)}^{\mathrm{m}}$$

where $v_i^{\mathrm{m}}$ is the nodal displacement at node $i$, $\varepsilon^{\mathrm{m}}$ is a linear map akin to an infinitesimal strain, $X_i$ is the (reference) nodal position and $\xi_{b(i)}^{\mathrm{m}}$ is a periodic correction. If $\varepsilon^{\mathrm{m}}$ is zero, the mechanism is *periodic* otherwise it is *affine*.

In a lattice with *affine mechanisms*, a macroscopic infinitesimal strain $\varepsilon(X) = \varepsilon^{\mathrm{m}}$ can be relaxed into a mechanism through the corresponding $\xi^{\mathrm{m}}$, making the effective strain energy density degenerate $\Phi^\star(\varepsilon^{\mathrm{m}}) = 0$. That does not imply, however, that a (compatible) *graded* strain, of the form $\varepsilon(X) = \lambda(X) \varepsilon^{\mathrm{m}}$ for some non-constant scalar field $\lambda$, can be relaxed into a mechanism as well. As a result, a significant mismatch can arise between the effective model which predicts zero energy, $\Phi^\star(\lambda(X) \varepsilon^m) = 0$, and the lattice model which predicts a non-zero energy. In that case, it is desirable to refine the effective model to capture the gradient effect, as in $\Phi^\star(\varepsilon, \nabla \varepsilon)$. Typical examples of degenerate effective strain energies regularized by strain gradient energies include pantographs [15, 18] and auxetic squares [45, 42]. The latter has been studied in the context of kirigami mechanics; see also the Miura origami pattern [36, 20] as well as other lattices proposed and referenced by Milton [31, 32]. Regularization can require including higher-order strain gradients; see [41]. It may also not be possible at all meaning that the lattice allows for the mechanism to be arbitrarily graded. Examples of degenerate effective strain energies, that cannot be regularized by strain gradients, include the square lattice as well as the "pentamode" materials useful in "cloaking" applications [37, 35, 34]. Note that, mechanisms and regularization aside, strain gradient elasticity also permits to model wave dispersion phenomena in crystals and metamaterials, chiral ones in particular; see, e.g., [39].

In a lattice with *periodic mechanisms*, the effective strain energy is not necessarily degenerate, because periodic mechanisms are averaged out and leave no macroscopic trace. Graded mechanisms, however, are still a source of concern. If a *gradient* of periodic mechanisms averages to a non-zero macroscopic strain, the effective strain energy will overestimate the lattice stiffness by not allowing the corresponding macroscopic strain to be relaxed by a gradient of periodic mechanisms. To remedy this, the effective strain energy density $\Phi^\star(\varepsilon)$ can be kinematically enriched into a density $\Phi^\star(\varepsilon, \nabla \theta)$ where $\theta$ is the amplitude of the periodic mechanism. The archetype of lattices with a periodic mechanism is the (fully extended) kagome lattice. Beyond it being featured in many computational designs [24], the kagome lattice has been investigated extensively in recent years in connection to the topological insulators of condensed matter physics [23]. In this trend, distorted kagome lattices have been shown to be polarized whereby they appear soft if indented on one side but stiff if indented on the opposite side [9]. Enriching the effective strain energy as described above is exactly what is needed to account for these polarization effects which, ultimately, are but a manifestation of localized, *i.e.*, graded, periodic mechanisms [33, 43]. Kinematically enriched elasticity, including but not limited to Cosserat's and micromorphic theories, has also been used to model bandgap crystals and metamaterials; see, e.g., [29, 7] and the work of Boutin and collaborators on inner-resonance phenomena in poroelastic media [10]. Other asymptotic methods, with emerging effective kinematics, have also been proposed to model wave phenomena at higher frequencies, *i.e.*, in the vicinity of simple or degenerate eigenfrequencies [13, 14]; see also $\boldsymbol{k} \cdot \boldsymbol{p}$ perturbation theory used in condensed matter physics [17].

In the present work, we propose a unified approach to lattice homogenization covering both the stable and unstable cases. The approach can in particular handle the *singular* cell problems that are encountered when periodic mechanisms are present, a case where it is natural to use the amplitude $\theta(\boldsymbol{X})$ of the periodic mechanism as an enrichment variable. We use second-order homogenization to resolve the effective energy in terms of both $\nabla\theta$, in the presence of periodic mechanisms, and $\nabla\boldsymbol{\varepsilon}$, which is helpful in the presence of affine mechanisms. We demonstrate the approach with particular lattices that illustrate a variety of effective behaviors. It follows a unified pattern and leads to a classification of these effective behaviors. The method at the core of our classification, later referred to as the "recipe", contains an asymptotic homogenization formalism that was derived and validated by the first two authors in a previous contribution dealing with stable lattices and frames [5]. Using formal two-scale expansions, asymptotic homogenization provides systematic rigorous macroscopic effective models for discrete and continuous microstructures [40, 6, 11]. The approach can be recast into a variational framework [8, 27, 2][5], using the microscopic elastic energy as a starting point and resulting in a compact formulation of the effective model. Here, the formalism introduced in [5] is modified to handle degeneracies due to the existence of mechanisms. The outcome, as stated above, is an effective strain energy regularized by gradients or with enriched kinematics and whose zeroes describe mechanisms, be them affine, periodic or graded. The work complements the approach of Abdoul-Anziz & Seppecher [1] who derived effective strain energies that are kinematically restricted to soft modes, i.e., to penalized mechanisms. It is also worthwhile to stress that none of the effective strain energies derived below are new per se: *the main novelty of the paper resides in clarifying the mechanistic microstructural origins of generalized effective strain energies of the strain-gradient or enriched types, in connection to the kind of available mechanisms.*

Lastly, it should be stressed that, in practice, there is no—experimental or numerical—difference between a mechanism on one hand and a deformation mode that has sufficiently low strain energy on the other hand. The existence of the latter, hereafter called a *soft* mode, should bear similar consequences on the effective strain energy. For instance, if the bonds are replaced with slender beams built-in at joints, then an unstable lattice becomes a stable frame that admits a set of soft modes. In some cases, the small yet non-zero energy of the soft modes is enough to regularize the effective strain energy (see, *e.g.*, the square lattice treated in [1]); in other cases, it is not, and gradient effects or kinematic enrichments should be considered, see, *e.g.*, the pantographic truss analyzed in [18]. Thus, in a sense, unstable lattices are an idealization of frames with soft modes (including "bending-dominant" frames [16]). The present paper adopts the idealized point of view of mechanisms for the most part and describes the sophistication needed to handle soft modes in its last section. But first, the paper begins with a summary of the main results and of the asymptotic approach developed in [5]. Relaxed assumptions on the stability of the microscopic model are introduced next, followed by a description of our classification procedure. The subsequent section focuses on the extension that allows the homogenization framework to handle mechanisms. The last two sections of the paper cover a series of examples (triangular lattice, square lattice, auxetic squares and kagome lattice) where the effective strain energy is computed and discussed in light of the proposed classification, first for mechanisms and then for soft modes.

## 2. OVERVIEW OF THE MAIN RESULTS

### 2.1. Homogenization method implemented in the shoal library: a summary

In this section, we summarize the variational asymptotic homogenization method introduced in [5]. This method has been implemented in a symbolic calculation language and is distributed as the open-source library shoal [3]. The results generated in Section 4 are based on an update of the library that handles lattices with periodic mechanisms, as discussed in Sections 2.2 and 3. The library is documented in full details in a technical memo that is included in the distribution.

The starting point of the homogenization procedure is a discrete network of "elastic connections" arranged into periodic cells of size $\gamma \ll 1$. The homogenization procedure can deal with arbitrary elastic connections in arbitrary dimension $d$, but we will focus here on a truss made out of elastic-linear *springs* in dimension $d = 2$. Each node in the lattice is assigned to a Bravais sub-lattice, defined as a group of nodes that are mapped to one another via the fundamental translations of the lattice. We denote the number of Bravais sub-lattices as $n_b$. The elastic lattice is viewed as a superposition of $n_b$ Bravais sub-lattices coupled by the energy of the springs. With this non-standard viewpoint, there is no need to define a unit cell—the information that is traditionally encoded in the unit cell is presented differently, see below. We denote as $n_\varphi$ the number of spring families, each family being a collection of springs that can be mapped to one another by the fundamental translations of the lattice. For instance, there are $n_\varphi = 3$ families of springs in the triangular lattice, and $n_\varphi = 2$ families in a square lattice.

Following the standard Cauchy-Born rule in homogenization, the nodal displacement

$$\boldsymbol{v}_i = \boldsymbol{u}(\boldsymbol{X}_i) + \gamma\, \boldsymbol{\xi}_{b(i)}(\boldsymbol{X}_i) \tag{2.1}$$

at any particular node $i$ in the lattice is parameterized using a set of smooth continuous functions: the macroscopic displacement $\boldsymbol{u}(\boldsymbol{X})$ and the microscopic shifts $\boldsymbol{\xi}_b(\boldsymbol{X})$. Therein, $b(i) \in \{I, \ldots, n_b\}$ denotes the Bravais sub-lattice of node $i$ and $\boldsymbol{X}_i$ its reference position. The discrete index $b(i)$ plays the role of the fast variable encountered in the homogenization of continuous microstructures.

Functions $\boldsymbol{u}(\boldsymbol{X})$ and $\boldsymbol{\xi}_b(\boldsymbol{X})$ are assumed to depend on $\gamma$ through regular power-series expansions, i.e., $\boldsymbol{u}(\boldsymbol{X}) = \sum_{k \geqslant 0} \gamma^k \boldsymbol{u}_{(k)}(\boldsymbol{X})$ and $\boldsymbol{\xi}_b(\boldsymbol{X}) = \sum_{k \geqslant 0} \gamma^k \boldsymbol{\xi}_{b,(k)}(\boldsymbol{X})$ where the series coefficient $\boldsymbol{u}_{(k)}(\boldsymbol{X})$ and $\boldsymbol{\xi}_{b,(k)}(\boldsymbol{X})$ are smooth functions of $\boldsymbol{X}$ that do not depend on $\gamma$. This regular dependence on $\gamma$ will however be implicit in our notation. We will assume that the gradients of the continuous fields with respect to the macroscopic variable $\boldsymbol{X}$ are all of order $\gamma^0$, i.e., $\nabla^k \boldsymbol{u} = \mathcal{O}(\gamma^0)$ and $\nabla^k \boldsymbol{\xi}_b = \mathcal{O}(\gamma^0)$ for any $k \geqslant 0$. This scaling law can be established formally using the series expansions—and is in fact the main concrete consequence of the existence of such expansions.

The microscopic shifts in (2.1) are chosen to satisfy the kinematical constraint

$$\forall \boldsymbol{X}, \quad \sum_{b \in \{I, \ldots, n_b\}} \boldsymbol{\xi}_b(\boldsymbol{X}) = \boldsymbol{0}. \tag{2.2}$$

By requiring that the average of the microscopic shifts over the different Bravais sub-lattices is zero, we make the macroscopic displacement $\boldsymbol{u}(\boldsymbol{X})$ the average of the continualized microscopic displacements over the Bravais sub-lattices, $\boldsymbol{u}(\boldsymbol{X}) = \frac{1}{n_b} \sum_{I \leqslant b \leqslant n_b} \left( \boldsymbol{u}(\boldsymbol{X}) + \gamma \boldsymbol{\xi}_b(\boldsymbol{X}) \right)$.

In terms of the nodal displacements $\boldsymbol{v}_i$, we define the linearized elongation in a spring $\alpha$ as

$$\varepsilon_\alpha = (\boldsymbol{v}_{i_\alpha^+} - \boldsymbol{v}_{i_\alpha^-}) \cdot \boldsymbol{t}_\alpha, \tag{2.3}$$

where $(i_\alpha^-, i_\alpha^+)$ are the endpoints of the spring, oriented in an arbitrary way, $\boldsymbol{t}_\alpha = (\boldsymbol{X}_{i_\alpha^+} - \boldsymbol{X}_{i_\alpha^-})/\ell_\alpha$ is the unit tangent in reference configuration, and $\ell_\alpha$ is the undeformed length. The elastic energy in the linearly elastic lattice can be written in terms of the spring elongations as

$$\Phi_{\text{d}} = \frac{1}{2} \sum_\alpha k \, \varepsilon_\alpha^2, \tag{2.4}$$

where $k$ denotes the spring constant, which we assume to be identical for all kinds of springs for the sake of simplicity. The spring elongation scales with the spatial period as $\varepsilon_\alpha = \mathcal{O}(\gamma)$, meaning that the energy $k \varepsilon_\alpha^2$ *per unit "volume"* scales as $k \gamma^{2-d}$, where $d$ is space dimension. This allows the infinite sum to converge in the continuous limit $\gamma \to 0$ as long as $k$ scales as $k = \mathcal{O}(\gamma^{d-2})$. In dimension $d = 2$ in particular, $k$ does not scale with $\gamma$. This is adopted and, from now on, we restrict theory to dimension 2.

The goal of asymptotic homogenization is precisely to identify a continuous approximation to the energy (2.4) in terms of the macroscopic strain and its gradients. Following Mandel's notation, we arrange the 3 independent components of macroscopic symmetric strain tensor $\boldsymbol{\varepsilon} = \frac{1}{2}(\nabla \boldsymbol{u} + \nabla \boldsymbol{u}^T)$ into a vector $\hat{\boldsymbol{\varepsilon}}$,

$$\hat{\boldsymbol{\varepsilon}}(\boldsymbol{X}) = \begin{pmatrix} \varepsilon_{11} & \varepsilon_{22} & \sqrt{2}\,\varepsilon_{12} \end{pmatrix}, \tag{2.5}$$

where the factor $\sqrt{2}$ applied to the shear strain is a convention used in the library, that needs to be recalled when interpreting its output.

We proceed to discuss the main steps in the homogenization procedure that are implemented in shoal.

**Continualized microscopic strain** The microscopic strain $\boldsymbol{E}(\boldsymbol{X})$ is a vector of length $n_E = n_\varphi + 2$, that collects the elongations $\overline{\varepsilon}_\varphi(\boldsymbol{X})$ relevant to the different spring families $\varphi \in \{1, \ldots, n_\varphi\}$ along with the components in the Cartesian basis of the vector appearing in the left-hand side of the constraint (2.2),

$$\boldsymbol{E}(\boldsymbol{X}) = \left( \overline{\varepsilon}_1(\boldsymbol{X}), \ldots, \overline{\varepsilon}_{n_\varphi}(\boldsymbol{X}), \sum_{I \leqslant b \leqslant n_b} (\boldsymbol{\xi}_b)_1(\boldsymbol{X}), \sum_{I \leqslant b \leqslant n_b} (\boldsymbol{\xi}_b)_2(\boldsymbol{X}) \right) \in \mathbb{R}^{n_E}. \tag{2.6}$$

The continualized spring elongation $\overline{\varepsilon}_\varphi(\boldsymbol{X})$ relevant to a particular spring family $\varphi$ is a symbolic expression defined as follows. For any spring $\alpha$ belonging to family $\varphi$, we insert the Cauchy-Born ansatz (2.1) for the nodal displacement in the elongation $\varepsilon_\alpha$ given by (2.3), and do a Taylor expansion of the nodal values $(\boldsymbol{u}(\boldsymbol{X}_{i_\alpha^-}), \boldsymbol{u}(\boldsymbol{X}_{i_\alpha^+}), \boldsymbol{\xi}_{b(i_\alpha^-)}(\boldsymbol{X}_{i_\alpha^-}),$ $\boldsymbol{\xi}_{b(i_\alpha^+)}(\boldsymbol{X}_{i_\alpha^+}))$ about the spring center $\boldsymbol{X}_c^\alpha = (\boldsymbol{X}_{i_\alpha^-} + \boldsymbol{X}_{i_\alpha^+})/2$. We identify the resulting expression as $\overline{\varepsilon}_\varphi(\boldsymbol{X}_c^\alpha)$. By construction, it is of the form $\overline{\varepsilon}_\varphi(\boldsymbol{X}_c^\alpha) = \overline{\boldsymbol{\varepsilon}}_y^\varphi \cdot \boldsymbol{y}(\boldsymbol{X}) + \gamma \overline{\boldsymbol{\varepsilon}}_y^{\varphi'} : \nabla \boldsymbol{y}(\boldsymbol{X}) + \cdots + \overline{\boldsymbol{\varepsilon}}_\varepsilon^\varphi \cdot \hat{\boldsymbol{\varepsilon}}(\boldsymbol{X}) + \gamma \overline{\boldsymbol{\varepsilon}}_\varepsilon^{\varphi'} : \nabla \hat{\boldsymbol{\varepsilon}}(\boldsymbol{X}) + \cdots$, where $\boldsymbol{y}(\boldsymbol{X})$ is the vector of microscopic degrees of freedom, obtained by concatenating the vectors $\boldsymbol{\xi}_I, \ldots, \boldsymbol{\xi}_{n_b}$,

$$\boldsymbol{y}(\boldsymbol{X}) = \begin{pmatrix} \boldsymbol{\xi}_I(\boldsymbol{X}) & \cdots & \boldsymbol{\xi}_{n_b}(\boldsymbol{X}) \end{pmatrix} \in \mathbb{R}^{n_y}, \qquad n_y = 2 n_b, \tag{2.7}$$

where the constant tensors $\overline{\boldsymbol{\varepsilon}}_y^\varphi, \ldots, \overline{\boldsymbol{\varepsilon}}_\varepsilon^{\varphi'}, \ldots$ depend on the lattice geometry and spring family, and are calculated in symbolic form by the library.

Denoting as $\mathcal{Q}$ the constant $2 \times n_E$ matrix defined in block-matrix notation by

$$\mathcal{Q} = \begin{pmatrix} \boldsymbol{0}_{2 \times n_\varphi} & \boldsymbol{I}_2 \end{pmatrix}, \tag{2.8}$$

where $\boldsymbol{I}_p$ is the identity matrix in dimension $p$, we rewrite the kinematical constraint (2.2) in a standard form expected by the library shoal as

$$\forall \boldsymbol{X}, \quad \mathcal{Q} \cdot \boldsymbol{E}(\boldsymbol{X}) = \boldsymbol{0}. \tag{2.9}$$

In the left-hand side, the matrix $\mathcal{Q}$ extracts the two trailing entries in $\boldsymbol{E}(\boldsymbol{X})$, see (2.6).

**Microscopic strain reconstruction** The microscopic strain $E(X)$ in (2.6) is known in terms of the microscopic degrees of freedom $y(X)$, of the macroscopic strain $\hat{\varepsilon}(X)$ and of their gradients: for the first $n_\varphi$ entries denoted $\bar{\varepsilon}_\varphi(X)$, this follows from the expansions described above, whereas for the last two entries in $E(X)$ representing the kinematical constraint, this is obvious from (2.6). We write this as $E(X) = E_{[\hat{\varepsilon},y]}(X)$ where

$$\begin{aligned} E_{[\hat{\varepsilon},y]}(X) = & \; E_\varepsilon \cdot \hat{\varepsilon}(X) + \gamma E'_\varepsilon : \nabla \hat{\varepsilon}(X) + \gamma^2 E''_\varepsilon \therefore \nabla^2 \hat{\varepsilon}(X) + \cdots \\ & E_y \cdot y(X) + \gamma E'_y : \nabla y(X) + \gamma^2 E''_y \therefore \nabla^2 y(X) + \cdots \end{aligned} \tag{2.10}$$

The constant tensors $E_\varepsilon$, $E'_\varepsilon$, $E''_\varepsilon$, $E_y$, $E'_y$ and $E''_y$ are calculated in symbolic form by the library shoal in terms of the lattice properties. Their dimensions are $(n_E, 3)$, $(n_E, 3, 2)$, $(n_E, 3, 2, 2)$, $(n_E, n_y)$, $(n_E, n_y, 2)$, and $(n_E, n_y, 2, 2)$ respectively.

**Continualized strain energy** Applying the Euler–macLaurin formula [25] to the discrete sum in the strain energy (2.4), we can rewrite it as an integral, see [5] for details,

$$\Phi[\hat{\varepsilon}, y] = \int_\Omega \frac{1}{2} E_{[\hat{\varepsilon},y]}(X) \cdot \mathcal{H} \cdot E_{[\hat{\varepsilon},y]}(X) \, \mathrm{d}X, \tag{2.11}$$

where $\mathcal{H} = \mathrm{diag}(k, \ldots, k, 0, 0)$. The first $n_\varphi$ entries in the diagonal of $\mathcal{H}$ are the spring constants of the different spring families. The two trailing zeroes play no role since the corresponding degrees of freedom are set to zero, see (2.6)–(2.9). In the right-hand side of (2.11), we have omitted the boundary terms that are produced by the Euler–macLaurin formula. Boundaries are consistently ignored in the present work, where we aim at deriving the effective properties in the bulk of the lattice. The kinematic constraint (2.9) is rewritten accordingly as

$$\forall X, \quad \mathcal{Q} \cdot E_{[\hat{\varepsilon},y]}(X) = 0. \tag{2.12}$$

**Macroscopic degrees of freedom** We define the vector $l(X)$ of macroscopic degrees of freedom by concatenating the 3 independent components of macroscopic strain in (2.5), with two sets of degrees of freedom $\theta(X)$ and $\rho(X)$ that are relevant to degenerate lattices:

$$l(X) = \begin{pmatrix} \hat{\varepsilon}(X) & \theta(X) & \rho(X) \end{pmatrix}. \tag{2.13}$$

As explained in Section 3.3, $\theta(X)$ collects the amplitudes of the periodic mechanisms (whenever present in the lattice) and $\rho(X)$ lists the Lagrange multipliers that enforce any macroscopic kinematic constraint (whenever applicable to the lattice). An important contribution of the present work is to rationalize the effective models produced by homogenization in terms of the additional degrees of freedom $\theta(X)$ and $\rho(X)$, corresponding to the two, and only two, possible ways in which the cell problem can become singular.

**Relaxation of the microscopic variables.** The homogenized energy $\Phi^\star$ is obtained by optimizing $\Phi[\hat{\varepsilon}, y]$ in (2.11) with respect to the microscopic variables $y(X)$, subjected to the constraint (2.12). In the spirit of formal asymptotic expansions, this relaxation step is performed order by order in the expansion parameter $\gamma$, and the library shoal delivers the solution for the microscopic shift in the form

$$y(X) = Y_0 \cdot l(X) + \gamma Y'_0 : \nabla l(X) + \mathcal{O}(\gamma^2). \tag{2.14}$$

The localization tensors $Y_0$, $Y'_0$ at the successive orders are obtained by solving unit cell problems. The sub-blocks in $Y_0$, $Y'_0$ corresponding to the degrees of freedom of the kind $\hat{\varepsilon}$ in $l$ capture the standard dependence of the microscopic shift $y$ on the macroscopic strain and its gradient, see (2.13). In lattices possessing mechanisms, the sub-blocks in $Y_0$, $Y'_0$ corresponding to $\theta$ capture an additional dependence on the amplitude of the mechanism and its gradient. The sub-blocks in $Y_0$ corresponding to $\rho$ are zero, which is not surprising as the degrees of freedom in $\rho$ can be interpreted as macroscopic stress variables.

**Details on leading-order homogenization** At dominant order $\gamma^0$, homogenization requires solving a cell problem in the form of a linear system

$$P \cdot \begin{pmatrix} y_0 \\ \lambda_0 \end{pmatrix} + R_0 \cdot \hat{\varepsilon} = 0. \tag{2.15}$$

The unknowns are the microscopic displacement $y_0$, as well as Lagrange multipliers $\lambda_0$ associated with the kinematical constraint (2.9). The problem (2.15) must be solved for any possible value of the macroscopic strain $\hat{\varepsilon} \in \mathbb{R}^3$. The symmetric square matrix $P$ and the rectangular matrix $R_0$ are calculated symbolically by the library shoal in terms of the lattice properties. The matrix $P$, in particular, is given by

$$P = \begin{pmatrix} \mathcal{W}_{yy} & (\mathcal{Q} \cdot E_y)^\top \\ \mathcal{Q} \cdot E_y & 0 \end{pmatrix}, \tag{2.16}$$

where

$$\mathcal{W}_{yy} = E_y^\top \cdot \mathcal{H} \cdot E_y. \tag{2.17}$$

In Section 3.3, we show that the solution to (2.15) is of the form $y_0 = Y_0 \cdot l$, in agreement with the leading-order term announced in (2.14). The relocalization tensor $Y_0$ is calculated in symbolic form by the library shoal. In the particular case of a stable lattice, there are neither $\theta$ nor $\rho$ degrees of freedom and $l$ in (2.13) coincides with $\hat{\varepsilon}$: in this case, $P$ in (2.15) can be inverted, yielding an expression of $y_0$ that is indeed linear in $\hat{\varepsilon} = l$. For unstable lattices, however, the solution $y_0 = Y_0 \cdot \begin{pmatrix} \hat{\varepsilon} & \theta & \rho \end{pmatrix}$ depends on the additional parameters $\theta$ and $\rho$, which can be interpreted as a set of coordinates in the null space of $P$, see Section 3.3 for details.

Inserting the leading-order solution $y_0 = Y_0 \cdot l$ for the microscopic shift into the energy (2.11), we obtain next the leading-order strain energy in the form $\int_\Omega K_0 : \frac{l \otimes l}{2} d^2 X$ where the homogenized stiffness $K_0$ is given by

$$K_0 = (E_y \cdot Y_0 + E_l)^\top \cdot \mathcal{H} \cdot (E_y \cdot Y_0 + E_l). \tag{2.18}$$

In the technical documentation of the library [3], we show that $K_0$ is zero on the degrees of freedom of the kind $\theta$ (amplitude of microscopic mechanisms) and $\rho$ (stress variables associated with macroscopic kinematic constraints). Denoting as $K_0^{\varepsilon\varepsilon}$ the sub-block of $K_0$ corresponding to the macroscopic strain, we can therefore rewrite the strain energy at leading order as

$$\int_\Omega K_0 : \frac{l \otimes l}{2} d^2 X = \int_\Omega K_0^{\varepsilon\varepsilon} : \frac{\hat{\varepsilon} \otimes \hat{\varepsilon}}{2} d^2 X. \tag{2.19}$$

The degrees of freedom $\theta$ and $\rho$ do enter the strain energy at higher orders, by a gradient effect.

**Homogenization to second order** The procedure is continued at the next orders $\gamma^1$ and $\gamma^2$. It is described in details in [5], and delivers the microscopic shift announced in (2.14). Inserting this solution $y(X)$ into (2.11), we obtain the effective energy $\Phi^\star[l]$ in the form of an expansion that depends on the macroscopic degrees of freedom $l(X)$ and their gradients,

$$\Phi^\star[l] = \int_\Omega \left( K_0 : \frac{l \otimes l}{2} + \gamma A_0 \therefore (l \otimes \nabla l) + \gamma^2 \left( B_0 :: \frac{\nabla l \otimes \nabla l}{2} + C_0 :: (l \otimes \nabla^2 l) \right) + \mathcal{O}(\gamma^3) \right) d^2 X, \tag{2.20}$$

where the orders of the various terms are indicated using the successive powers $\gamma^i$ of the scale separation parameter $\gamma$. The functional $\Phi^\star$ is asymptotically exact up to boundary terms that are calculated by the library but have been ignored here, see [5]. The detailed derivation of the operators $Y_0'$, $A_0$, $B_0$ and $C_0$ can be found in [5].

## 2.2. Assumption on microscopic stability

We exclude material instabilities in the microscopic model, by assuming that the microscopic elasticity matrix $\mathcal{H}$ is positive-definite on kinematically admissible microscopic strain $E$, that is

$$(\forall E \text{ s.t. } Q \cdot E = 0), \quad \begin{cases} \frac{1}{2} E \cdot \mathcal{H} \cdot E \geqslant 0 \\ E \neq 0 \Rightarrow \frac{1}{2} E \cdot \mathcal{H} \cdot E > 0 \end{cases} \tag{2.21}$$

In particular, for a microscopic strain produced purely by a microscopic shift, i.e. for $E = E_y \cdot y$, this implies

$$(\forall y \text{ s.t. } Q \cdot E_y \cdot y = 0), \quad \frac{1}{2} y \cdot \mathcal{W}_{yy} \cdot y = 0 \Rightarrow E_y \cdot y = 0, \tag{2.22}$$

where $\mathcal{W}_{yy}$ has been defined in (2.17).

The assumption (2.21) is weaker than that used in our earlier work [5], where we excluded periodic mechanisms by assuming instead ($\forall y$ such that $Q \cdot E_y \cdot y = 0$), $\frac{1}{2} y \cdot \mathcal{W}_{yy} \cdot y = 0 \Rightarrow y = 0$. Periodic mechanisms are such that $E_y \cdot y = 0$ and $y \neq 0$, see Section 3.2.

## 2.3. Classification tree

In this section, we present the sequence of steps involved in the analysis of a lattice. The sequence is represented as a tree in Figure 2.1, which helps classifying the different types of effective behaviors.

1. For a given microstructure defined by a discrete energy in the form (2.4), we rely on the shoal library [3] to work out the constrained energy formulation in (2.10)–(2.12).
2. The library assembles the symmetric matrix $P$ that enters in the unit cell problems (2.15).
3. It checks that the rank deficiency of $P$ is consistent with the integer value rankDeficiency declared by the user,
   - if the values are inconsistent, information about the null space of $P$ is printed out and the procedure terminates with a message suggesting to run again the homogenization with a corrected value of rankDeficiency.
4. If $P$ is singular, the library computes the null space of $P$ which reveals:
   - applicable macroscopic kinematic constraints, if any, see Section 3.1,
   - microscopic, periodic mechanisms, if any, see Section 3.2.
5. The library calculates the matrix in the right-hand side of (3.10) furnishing the general solution to the leading-order cell problem, as well as the leading-order stiffness $K_0$ using (2.18), and extracts its sub-block $K_0^{\varepsilon\varepsilon}$, see (2.19).
6. If there are no microscopic periodic mechanisms $\theta$ and the sub-block $K_0^{\varepsilon\varepsilon}$ is full-rank, the effective elasticity is non-degenerate at leading order, and the homogenization is complete.
7. Otherwise, higher-order homogenization is required, because of the presence of periodic mechanisms (parameterized by $\theta$) and/or because $K_0^{\varepsilon\varepsilon}$ is not full-rank. The library then derives the order-$\gamma^2$ effective energy $\Phi^\star$ in (2.20). It is then up to the user to analyze the higher-order energy and assess whether it accurately captures all the modes of deformation that can be present in the original lattice.

**Remark 2.1.** The procedure can only identify microscopic periodic mechanisms whose spatial period matches the period of the lattice. To capture periodic mechanisms whose spatial period is a *multiple* of the period of the lattice [22, 26], a Bloch analysis [19] needs to be performed prior to running the homogenization procedure. If a periodic mechanism extending over multiple unit cells is identified, the homogenization procedure should be set up on an 'extended' cell containing several replica of the unit cell.

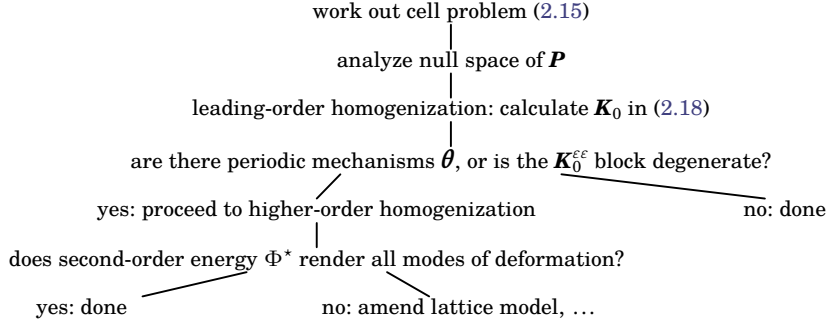

**Figure 2.1.** Decision tree illustrating graphically the analysis of a particular lattice. The ellipsis at the bottom of the tree indicates that alternative fixes may be identified in some cases, such as pushing the homogenization to even higher orders.

We further illustrate this procedure on a set of examples that are summarized the Table 2.1 below and analyzed in detail in Section 4.

| | rank defic. of $P$ | null vectors of $P$ | null vectors of $K_0^{\varepsilon\varepsilon}$ | final model |
|---|---|---|---|---|
| triangular | 0 | none | none | (4.1): classical elasticity |
| square | 0 | none | shear | (4.3): persistent shear mechanism |
| auxetic squares (generic) | 0 | none | dilation | (4.9): mixed classical (deviatoric) and strain-gradient elasticity (dilation) |
| auxetic squares (fully extended) | 1 | $\theta$ (twist) | none | (4.19): uncoupled classical elasticity and gradient of enrichment variable |
| auxetic squares (with alignment) | 0 | none | dilation | (4.20): persistent dilation mechanism |
| kagome (generic) | 0 | none | dilation | (4.28): persistent dilation mechanism |
| kagome (fully extended) | 1 | $\theta$ (twist) | none | (4.34): coupled classical elasticity and gradient of enrichment variable |
| honeycomb made up of inext. beams | 1 | $\rho$ (incompr.) | regular | Eq. [5.9] from [44]: classical, 2D-incompressible elasticity |

**Table 2.1.** Application of the method of analysis to various lattices: summary of the main findings. All lattices but the last one are truss lattices that are analyzed in detail in Section 4. The honeycomb lattice made out of inextensible beams (last row in the table) is analyzed in [44] using the shoal library, and illustrates a rank deficiency caused by a macroscopic kinematic constraint (Section 3.1).

## 3. ANALYSIS OF A RANK-DEFICIENT UNIT CELL PROBLEM

In this section, we investigate the case where the square matrix $P$ entering in the cell problem (2.15) is rank-deficient, relevant to lattices having periodic mechanisms. The practical method we obtain has been implemented in an extension of the library [5], see Section E of the technical memo for details.

Assuming that the matrix $P$ is singular, we denote as $(y^\star, \lambda^\star) \neq 0$ a null vector, such that

$$P \cdot \begin{pmatrix} y^\star \\ \lambda^\star \end{pmatrix} = 0. \tag{3.1}$$

The matrix $P$ in (2.16) is square and symmetric, and this is both a left- and a right- null vector. We will start by discussing the case treated in the original version [5], that is when $y^\star = 0$ (Section 3.1) and then extend to the case when $y^\star \neq 0$ (Section 3.2).

### 3.1. Case 1: macroscopic kinematic constraint

To start with, we consider the special case where the null vector is of the form $(y^\star, \lambda^\star) = (0, \lambda^\star)$, with $\lambda^\star \neq 0$, *i.e.*,

$$P \cdot (0, \lambda^\star) = 0. \tag{3.2}$$

Equations (2.16) and (3.1) can then be rewritten as

$$\lambda^\star \cdot Q \cdot E_y = 0, \tag{3.3}$$

that is $\lambda^\star \in \ker((Q \cdot E_y)^\top)$. Let us combine linearly the kinematic constraints appearing in (2.9) with coefficients given by the entries of $\lambda^\star$, namely $0 = \lambda^\star \cdot Q \cdot E = \lambda^\star \cdot Q \cdot (E_y \cdot y + E_l \cdot l) = [\lambda^\star \cdot Q \cdot E_l] \cdot l$. This particular combination of the kinematic constraints depends only on the *macroscopic* degrees of freedom $l$,

$$[\lambda^\star \cdot Q \cdot E_l] \cdot l = 0. \tag{3.4}$$

As long as $[\boldsymbol{\lambda}^\star \cdot \mathcal{Q} \cdot \boldsymbol{E}_l] \neq \boldsymbol{0}$, Equation (3.4) will be referred to as a *macroscopic* kinematic constraint.

## 3.2. Case 2: microscopic, *periodic* mechanisms

Next, consider the case where the null vector $(\boldsymbol{y}^\star, \boldsymbol{\lambda}^\star)$ of $\boldsymbol{P}$ is such that $\boldsymbol{y}^\star \neq \boldsymbol{0}$. Equations (2.16) and (3.1) can then be rewritten as

$$\begin{aligned} \mathcal{W}_{yy} \cdot \boldsymbol{y}^\star + \boldsymbol{\lambda}^\star \cdot \mathcal{Q} \cdot \boldsymbol{E}_y &= \boldsymbol{0} \\ \mathcal{Q} \cdot \boldsymbol{E}_y \cdot \boldsymbol{y}^\star &= \boldsymbol{0} \end{aligned} \quad (3.5)$$

Multiplying the first equation by $\boldsymbol{y}^\star$ and using the second equation, we get

$$\boldsymbol{y}^\star \cdot \mathcal{W}_{yy} \cdot \boldsymbol{y}^\star = \boldsymbol{0} \quad \text{and} \quad \mathcal{Q} \cdot \boldsymbol{E}_y \cdot \boldsymbol{y}^\star = \boldsymbol{0}. \quad (3.6)$$

Using the property (2.22), we conclude that $\boldsymbol{y}^\star$ is a periodic mechanism,

$$\boldsymbol{E}_y \cdot \boldsymbol{y}^\star = \boldsymbol{0}. \quad (3.7)$$

Multiplying by $\boldsymbol{E}_y^T \cdot \mathcal{H}$ on the left-hand side, we further have $\mathcal{W}_{yy} \cdot \boldsymbol{y}^\star = \boldsymbol{0}$, which together with $(3.5)_2$ shows that the *projection* $(\boldsymbol{y}^\star, \boldsymbol{0})$ of the null vector onto the microscopic degrees of freedom is a null vector of $\boldsymbol{P}$ on its own,

$$\boldsymbol{P} \cdot (\boldsymbol{y}^\star, \boldsymbol{0}) = \boldsymbol{0}. \quad (3.8)$$

The null vector $(\boldsymbol{y}^\star, \boldsymbol{0})$ describes a *microscopic periodic mechanism*. The *twisting* of triangles in the fully extended kagome lattice and of squares in the fully extended auxetic square lattice are examples of microscopic periodic mechanisms, and are discussed in Sections 4.4 and 4.7 respectively.

The projection of the original null vector $(\boldsymbol{y}^\star, \boldsymbol{\lambda}^\star)$ onto the space of Lagrange multipliers, $(\boldsymbol{0}, \boldsymbol{\lambda}^\star) = (\boldsymbol{y}^\star, \boldsymbol{\lambda}^\star) - (\boldsymbol{y}^\star, \boldsymbol{0})$, is another null vector of $\boldsymbol{P}$, as can be shown easily, and it is of the first kind, analyzed in Section 3.1. It is therefore sufficient to analyze the null vectors of form $(\boldsymbol{y}^\star, \boldsymbol{0})$ on the one hand, and of the form $(\boldsymbol{0}, \boldsymbol{\lambda}^\star)$ on the other hand.

## 3.3. Solution to the singular cell problem

For any singular matrix $\boldsymbol{P}$, the general solution to the leading-order cell problem (2.15) can be written as a particular solution given by a pseudo-inverse $\boldsymbol{P}^{-1}$, plus a linear combination of the null vectors,

$$\begin{pmatrix} \boldsymbol{y}_0 \\ \boldsymbol{\lambda}_0 \end{pmatrix} = -\boldsymbol{P}^{-1} \cdot \boldsymbol{R}_0 \cdot \hat{\boldsymbol{\varepsilon}} + \boldsymbol{N}_y^\top \cdot \boldsymbol{\theta} + \boldsymbol{N}_\lambda^\top \cdot \boldsymbol{\rho}, \quad (3.9)$$

where $\boldsymbol{\theta} = (\theta_1, \theta_2, \dots)$ is a vector made out of the amplitudes of the microscopic periodic mechanisms, $\boldsymbol{\rho} = (\rho_1, \rho_2, \dots)$ is the vector of Lagrange multipliers dual to the macroscopic kinematic constraints, and $\boldsymbol{N}_y$ and $\boldsymbol{N}_\lambda$ are constant matrices that store the null vectors of $\boldsymbol{P}$ of either kind,

$$\begin{aligned} \boldsymbol{N}_y &= \begin{pmatrix} \boldsymbol{y}_1^\star & \boldsymbol{0} \\ \boldsymbol{y}_2^\star & \boldsymbol{0} \\ \vdots & \vdots \end{pmatrix} \\ \boldsymbol{N}_\lambda &= \begin{pmatrix} \boldsymbol{0} & \boldsymbol{\lambda}_1^\star \\ \boldsymbol{0} & \boldsymbol{\lambda}_2^\star \\ \vdots & \vdots \end{pmatrix} \end{aligned}.$$

By using the vector $\boldsymbol{l} = (\hat{\boldsymbol{\varepsilon}}, \boldsymbol{\theta}, \boldsymbol{\rho})$ of macroscopic degrees of freedom introduced in (2.13), one can rewrite (3.9) as

$$\begin{pmatrix} \boldsymbol{y}_0 \\ \boldsymbol{\lambda}_0 \end{pmatrix} = \left( -\boldsymbol{P}^{-1} \cdot \boldsymbol{R}_0 \cdot (\boldsymbol{I}, \boldsymbol{0}, \boldsymbol{0}) + \boldsymbol{N}_y^\top \cdot (\boldsymbol{0}, \boldsymbol{I}, \boldsymbol{0}) + \boldsymbol{N}_\lambda^\top \cdot (\boldsymbol{0}, \boldsymbol{0}, \boldsymbol{I}) \right) \cdot \boldsymbol{l}. \quad (3.10)$$

We denote as $\boldsymbol{Y}_0$ the upper block of the matrix appearing in the right-hand side. It delivers the leading-order microscopic shift as $\boldsymbol{y}_0 = \boldsymbol{Y}_0 \cdot \boldsymbol{l}$, as announced in (2.14).

The case of a stable lattice and a regular matrix $\boldsymbol{P}$ can be viewed as a particular case: the pseudo-inverse $\boldsymbol{P}^{-1}$ then becomes the unique inverse of $\boldsymbol{P}$ and the degrees of freedom $\boldsymbol{\theta}$ and $\boldsymbol{\rho}$ disappear.

The above solution method can be easily adapted to deal with the higher-orders cell problem, which use the same matrix $\boldsymbol{P}$ as the leading-order (2.15) but a different source term than $\boldsymbol{R}_0 \cdot \hat{\boldsymbol{\varepsilon}}$, see Appendix E of the technical documentation distributed with the library [3].

Null vectors of the first kind (macroscopic kinematic constraints) have already been discussed in our previous work: the reader is referred to the analysis of the honeycomb lattice made out of inextensible beam elements based on the shoal library, which appeared in Section 5.4 of [44] and is summarized in the last row in Table 2.1. The macroscopic kinematic constraint (3.4) takes the form of an area-preservation constraint $\operatorname{tr} \boldsymbol{\varepsilon} = 0$, see Eq. [5.9] in [44]. The associated macroscopic degree of freedom $\rho$ is a stress variable, which can be interpreted at the macroscopic level as a pressure-like Lagrange multiplier enforcing 2D incompressibility, and at the microscopic level as a compressive normal force distributed in all the beams of the lattice.

The fully extended auxetic squares analyzed in Section 4.4 is an example of an unstable lattice whose singular matrix $\boldsymbol{P}$ has a null vector of the second kind, see the microscopic periodic mechanism in (4.13). Following the general strategy outlined above, the amplitude $\theta$ of the mechanism is treated as an enrichment variable and included in the list $\boldsymbol{l}$ of degrees of freedom, see (2.13) and (4.14). The gradient of the enrichment variable $\theta$ appears in the final energy (4.19).

## 4. EXAMPLES

We apply the proposed method to a selection of lattices and derive the effective strain energy in closed form. Regularization and enrichment are discussed in light of the available mechanisms. The code which was used to generate the forthcoming results is written in the symbolic language Wolfram Mathematica and is included in the latest version of the shoal library [4].

In all the examples, the scale separation parameter $\gamma$ is the ratio $\gamma = \ell/L$ of the microscopic size $\ell$ (which is the undeformed spring length in most examples) to a macroscopic size $L$. We use the convention that the macroscopic size $L = \mathcal{O}(1)$ is independent of $\gamma$. This allows to use $\ell$ in lieu of the expansion parameter $\gamma$ as $\mathcal{O}(\ell) = \mathcal{O}(\gamma)$.

### 4.1. Triangular lattice

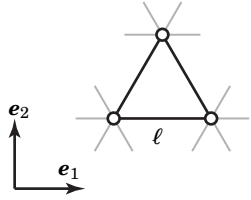

**Figure 4.1.** Unit cell of a triangular lattice.

The triangular lattice (Figure 4.1) is a simple lattice, *i.e.* all its nodes are equivalent ($n_b = 1$). There is no microscopic degrees of freedom in simple lattices and the unit cell problem (2.15) is trivial, *i.e.*, it degenerates to $\mathbf{0} = \mathbf{0}$. The effective strain energy (2.20) is delivered by the library at leading order $\gamma^0$ in the form

$$\begin{aligned}\Phi^\star_{[0]} &= \iint \frac{1}{2}(\lambda\,\text{tr}^2\boldsymbol{\varepsilon} + 2\mu\,\boldsymbol{\varepsilon}:\boldsymbol{\varepsilon})\,\mathrm{d}^2\mathbf{X} \\ &= \iint \frac{1}{2}((\lambda+\mu)\,\text{tr}^2\boldsymbol{\varepsilon} + 2\mu\,\boldsymbol{\varepsilon}^D:\boldsymbol{\varepsilon}^D)\,\mathrm{d}^2\mathbf{X},\end{aligned} \qquad (4.1)$$

where $\boldsymbol{\varepsilon}:\boldsymbol{\varepsilon} = \text{tr}(\boldsymbol{\varepsilon}\cdot\boldsymbol{\varepsilon})$ denotes the contraction with respect to a pair of indices, and $\boldsymbol{\varepsilon}^D = \boldsymbol{\varepsilon} - \frac{1}{2}\text{tr}\,\boldsymbol{\varepsilon}\,\mathbf{I}_2$ denotes the deviatoric part of the strain tensor in dimension $d = 2$. The Lamé coefficients are found as

$$\lambda = \mu = \frac{\sqrt{3}\,k}{4}. \qquad (4.2)$$

The effective medium is an isotropic Cauchy medium and is non-degenerate since $\lambda + \mu > 0$ and $\mu > 0$. Stated differently, the leading-order effective elasticity $\mathbf{K}_0 = \mathbf{K}_0^{\varepsilon\varepsilon}$ is positive definite.

Regularization and enrichment are not needed and, if rather forcefully introduced, they would provide asymptotically negligible corrections. These results are typical of stable lattices.

### 4.2. Square lattice

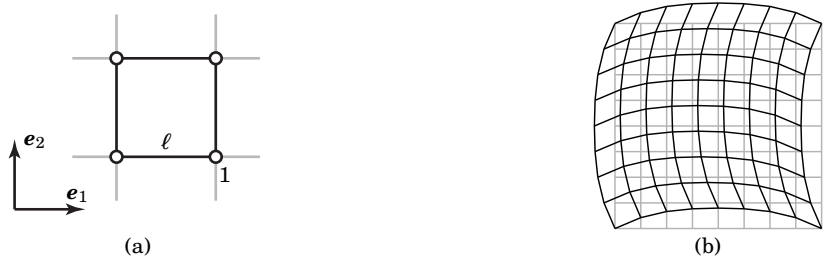

**Figure 4.2.** The square lattice: (a) a unit cell; (b) a mechanism with parabolic profiles.

The square lattice is yet another simple lattice (Figure 4.2a). The cell problem is again trivial and homogenization yields, at leading order,

$$\Phi^\star_{[0]} = \iint \frac{k}{2}(\varepsilon_{11}^2 + \varepsilon_{22}^2)\,\mathrm{d}^2\mathbf{X}. \qquad (4.3)$$

As earlier in (4.1), we denote as $\Phi^\star_{[k]}$ the terms of order $\gamma^k$ in the effective energy (2.20).

The effective strain energy density $\Phi^\star_{[0]}$ in (4.3) is anisotropic and, more importantly, degenerate relative to $\varepsilon_{12}$. Stated differently, the leading-order effective elasticity $\mathbf{K}_0 = \mathbf{K}_0^{\varepsilon\varepsilon}$ is rank-deficient, with a null space generated by the shear mode. Macroscopic displacements of the form $\mathbf{u}(\mathbf{X}) = f(X_2)\,\mathbf{e}_1 + g(X_1)\,\mathbf{e}_2$, where $f$ and $g$ are arbitrary functions, have $\varepsilon_{11} \equiv 0$ and $\varepsilon_{22} \equiv 0$, and therefore $\Phi^\star_{[0]} = 0$: they are macroscopic mechanisms. These mechanisms are depicted in Figure 4.2(b).

These mechanisms are not a homogenization shortcoming and higher-order corrections will not be able to assign them a non-zero strain energy. Indeed, pushing the asymptotics of the effective strain energy to higher orders delivers a *vanishing* correction at order $\gamma^1$ because of centrosymmetry,

$$\Phi^\star_{[1]} = 0, \qquad (4.4)$$

and a correction at order $\gamma^2$ which does not penalize $\varepsilon_{12}$, as anticipated,

$$\Phi^\star_{[2]} = \iint \frac{k\,\ell^2}{24} \left( \varepsilon_{11} \frac{\partial^2 \varepsilon_{11}}{\partial X_1^2} + \varepsilon_{22} \frac{\partial^2 \varepsilon_{22}}{\partial X_2^2} \right) d^2\boldsymbol{X}. \quad (4.5)$$

A natural way of penalizing these pure-shear mechanisms and regularizing the strain energy is to amend the discrete model, *e.g.*, by replacing pin joints with built-in joints, see [1].

## 4.3. Auxetic squares (generic case)

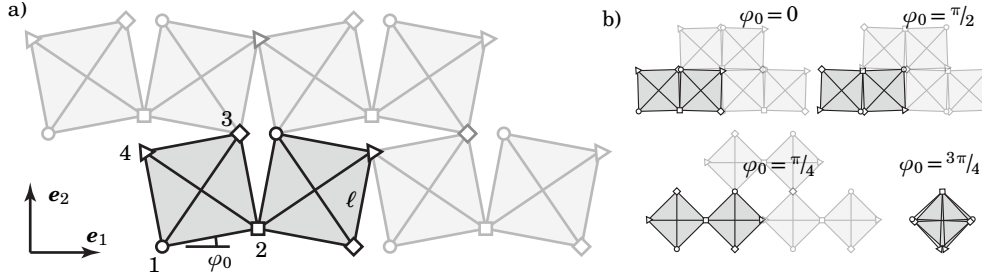

**Figure 4.3.** A reference configuration of the auxetic squares lattice, parameterized by a pre-twist angle $\varphi_0$: (a) generic value of $\varphi_0$, (b) special values: close packed ($\varphi_0 = 0, \pi/2$), fully extended ($\varphi_0 = \pi/4$) and collapsed ($\varphi_0 = 3\pi/4$). The lattice is made up of springs (black lines) connecting nodes (symbols). There are $n_b = 4$ Bravais sub-lattices, as indicated by the different symbols for the nodes. The grey color only serves to highlight the blocks that deform rigidly when the mechanism is activated.

|  | $\varphi_0$ | $\boldsymbol{P}$ | $b$ in (4.8) |
|---:|:---:|:---:|:---:|
| close packed | $0$ | regular | $=0$ |
| fully extended | $\pi/4$ | singular | $>0$ |
| close packed | $\pi/2$ | regular | $=0$ |
| collapsed (excluded) | $3\pi/4$ | divergent |  |

**Table 4.1.** Special initial configurations of the auxetic square lattice, and main features of their upcoming analyses.

The auxetic square lattice is illustrated on Figure 4.3(a) in a generic state where the squares are pre-twisted through an angle $\varphi_0$ chosen in the range $0 \leqslant \varphi_0 < \pi$. Incrementing $\varphi_0$ by $\pi$ amounts to rotate the lattice rigidly by an angle $\pi$, so that the proposed range $\varphi_0 \in [0, \pi)$ covers all possible configurations. In the course of the analysis, we will encounter four particular values of $\varphi_0$ which need special attention. They are announced in Table 4.1. In particular, we exclude from the analysis the case $\varphi_0 = 3\pi/4$ shown in Figure 4.3(b), where the entire lattice shrinks to a square: in this peculiar configuration, the macroscopic coordinate $\boldsymbol{X}$ cannot be defined and the two-scale analysis is inapplicable. Note that the springs are allowed to cross and overlap, as happens for instance in the configurations $\varphi_0 = 0, \pi/2$ and $3\pi/4$ in Figure 4.3(b).

Applying the homogenization "recipe" with a symbolic value of $\varphi_0$, we find that the matrix $\boldsymbol{P}$ governing the cell problem is regular, except for $\varphi_0 = \pi/4$ (and for $\varphi_0 = 3\pi/4$, a case which we have already excluded). The singular value $\varphi_0 = \pi/4$ corresponds to a fully extended initial configuration, see Figure 4.3(b), and is analyzed later in Section 4.4. For the moment, we focus on the case $\varphi_0 \notin \{\pi/4, 3\pi/4\}$ where $\boldsymbol{P}$ is regular and the only macroscopic degrees of freedom are therefore the macroscopic strain $\boldsymbol{l} = \hat{\boldsymbol{\varepsilon}}$.

The leading-order strain energy density is

$$\Phi^\star_{[0]} = \frac{1}{2} \iint (2\mu)\, \boldsymbol{\varepsilon}^D : \boldsymbol{\varepsilon}^D \, d^2\boldsymbol{X}, \quad (4.6)$$

where the effective shear modulus is given by $\mu = \frac{k}{2}$. The effective energy $\Phi^\star_{[0]}$ depends only on the deviatoric strain $\boldsymbol{\varepsilon}^D$, and is thus degenerate with respect to dilations, $\boldsymbol{\varepsilon}^m = \boldsymbol{I}_2$, which corresponds to the twisting mechanism. Unfortunately, the effective model accommodates many more mechanisms, namely all conformal transformations of the form $\boldsymbol{\varepsilon} = \lambda(\boldsymbol{X})\,\boldsymbol{I}_2$, where $\lambda$ is not constant in space and such that the strain $\boldsymbol{\varepsilon}$ is geometrically compatible. This is a shortcoming of the leading-order effective strain energy because the lattice only accommodates the *non-graded* version of the mechanism ($\nabla \lambda = \boldsymbol{0}$).

To remedy this, higher-order correctors must be taken into account. Due to the centro-symmetry of the lattice, the order-$\ell$ correction in the effective energy vanishes, $\Phi^\star_{[1]} = 0$. At order $\ell^2$, we obtain

$$\Phi^\star_{[2]} = \iint \frac{k\,\ell^2}{2} \left( \begin{array}{l} b(\varphi_0)\,\|\nabla \operatorname{tr} \boldsymbol{\varepsilon}\|^2 + (\cdots)\,\nabla \boldsymbol{\varepsilon}^D \otimes \nabla \operatorname{tr} \boldsymbol{\varepsilon} + (\cdots)\,\nabla \boldsymbol{\varepsilon}^D \otimes \nabla \boldsymbol{\varepsilon}^D \\ + (\cdots)\,\boldsymbol{\varepsilon}^D \otimes \nabla^2 \operatorname{tr} \boldsymbol{\varepsilon} + (\cdots)\,\boldsymbol{\varepsilon}^D \otimes \nabla^2 \boldsymbol{\varepsilon}^D \end{array} \right) d^2\boldsymbol{X}, \quad (4.7)$$

where the coefficient of the $\|\nabla \operatorname{tr} \boldsymbol{\varepsilon}\|^2$ term is

$$b(\varphi_0) = \frac{\cos^2 \varphi_0 \sin^2 \varphi_0}{2(1 - 2\cos\varphi_0 \sin\varphi_0)} \quad \begin{cases} = 0 & \text{for } \varphi_0 \in \{0, \pi/2\}, \\ > 0 & \text{otherwise.} \end{cases} \quad (4.8)$$

As shown in Table 4.1, a cancellation $b(\varphi_0) = 0$ takes place for $\varphi_0 = 0$ and $\varphi_0 = \pi/2$. This corresponds to a special configuration of the undeformed lattice, analyzed later in Section 4.5, where some of the springs are aligned and the squares are close-packed. We leave the analysis of this special case for later and focus on the case $b(\varphi_0) > 0$.

Although all terms in $\Phi^\star_{[2]}$ contribute equally, a priori, see (4.7), only the $b\,\|\nabla\,\mathrm{tr}\,\varepsilon\|^2$ term is appropriate to keep. Indeed, the other ones involve gradients $\nabla \varepsilon^D$ and $\nabla^2 \varepsilon^D$ of the deviatoric strain that are dominated by contributions from $\varepsilon^D$ to $\Phi^\star_{[0]}$.

Accordingly, an appropriate effective strain energy is

$$\Phi^\star = \frac{1}{2}\iint \left(2\mu\,\varepsilon^D:\varepsilon^D + k\,\ell^2\,b(\varphi_0)\,\|\nabla\,\mathrm{tr}\,\varepsilon\|^2\right)\,\mathrm{d}^2 X. \tag{4.9}$$

This is an isotropic semi-definite energy of the strain gradient type that vanishes exclusively over *uniform* macroscopic dilations ($\mathrm{tr}\,\varepsilon \equiv \mathrm{cst}$, $\varepsilon^D \equiv 0$). Thanks to the gradient correction, conformal transformations are no longer mechanisms and are instead soft modes with an energy that scales like $\ell^2$. Conversely, the correction is most important to keep in regimes governed by soft modes defined by $\varepsilon^D \sim \ell\,\nabla\,\mathrm{tr}\,\varepsilon$. If the deviatoric part is any larger, it dominates and the correction becomes superfluous (except perhaps from a numerical standpoint).

In anticipation of Section 4.4, it is useful to note that the leading-order microscopic shift $y_0 = Y_0(\varphi_0)\cdot\hat\varepsilon$ underpinning (4.6) is made out of a contribution coupled to the macroscopic dilation $\mathrm{tr}\,\varepsilon$, that diverges when $\varphi_0 \to \pi/4$, and a contribution coupled to $\varepsilon^D$ that remains bounded (see the companion notebook [4]),

$$y_0(\varepsilon) = \underbrace{\left(\frac{1}{1-\tan\varphi_0} - \frac{1}{2}\right)\mathrm{tr}\,\varepsilon}_{\text{twist angle}} \times y^\theta(\varphi_0) + (\cdots)\,\varepsilon^D. \tag{4.10}$$

The vector $y^\theta(\varphi_0)$ associated with the dilation mode describes square twisting, whereby the adjacent squares rendered in grey in Figure 4.3 rotate *rigidly* in alternating directions according to

$$y^\theta(\varphi_0) = \frac{1}{\cos\varphi_0 + \sin\varphi_0}(\ +e_1\ \ +e_2\ \ -e_1\ \ -e_2\ ), \tag{4.11}$$

see (2.7). The described twisting motion has been normalized in such a way that the squares rotate by an angle $\pm 1$ when $y = y^\theta(\varphi_0)$. As a result, the coefficient $\left(1/_{1-\tan\varphi_0} - 1/_2\right)\mathrm{tr}\,\varepsilon$ of $y^\theta(\varphi_0)$ in (4.10) can be interpreted as the twisting angle, i.e., the angle of rotation of one of the squares. We refer the reader to the companion Mathematica notebook for the detailed proofs [4]. One can now foresee the changes needed in the theory: as $\varphi_0$ approaches $\pi/4$, the nodal displacements in (4.10) diverge unless $\mathrm{tr}\,\varepsilon$ approached 0. In that case, the twisting amplitude becomes indeterminate. At the exact value $\varphi_0 = \pi/4$, the twisting mode assumes a life of its own, independent of $\varepsilon$ and its trace, and the twisting amplitude, hereafter denoted $\theta$, needs be treated as an additional kinematic variable. This is done next.

## 4.4. Special case of fully extended auxetic squares

Thus far, all encountered unit cell problems have been regular (i.e., matrix $P$ is invertible). However, for a fully extended lattice of auxetic squares,

$$\varphi_0 = \frac{\pi}{4}, \tag{4.12}$$

the unit cell problem is singular and matrix $P$ has a kernel of dimension 1; see equation (3.8). The generator of the kernel is the twisting mode, already encountered in (4.11),

$$y^\theta(\pi/4) = \frac{1}{\sqrt{2}}(\ +e_1\ \ +e_2\ \ -e_1\ \ -e_2\ ). \tag{4.13}$$

By (2.1) (with $\gamma = \ell$) and (2.7), it generates a microscopic displacement $\ell\,\xi_I = \ell\,e_1/\sqrt{2}$, $\ell\,\xi_{II} = \ell\,e_2/\sqrt{2}$, $\ell\,\xi_{III} = -\ell\,e_1/\sqrt{2}$ and $\ell\,\xi_{IV} = -\ell\,e_2/\sqrt{2}$ (on each of the Bravais lattices) that makes the grey squares in Figure 4.3 rotate rigidly with alternating angles $\pm 1$.

Back to the homogenization "recipe" presented earlier, the macroscopic strain measures gathered in the vector $l$ are automatically extended to include a fourth parameter representing the indeterminate amplitude of the microscopic mechanism, see (2.13) and (2.5)

$$l = \Big(\underbrace{\varepsilon_{11}, \varepsilon_{22}, \sqrt{2}\,\varepsilon_{12}}_{\varepsilon}, \theta\Big). \tag{4.14}$$

The fourth parameter $\theta$ has been normalized to represent the twisting angle, such that the leading-order microscopic shift is

$$y_0 = y_0(l(X)) = \theta(X)\,y^\theta(\pi/4) + (\cdots)\,\varepsilon^D(X). \tag{4.15}$$

The similarity with the expression (4.10) valid for $\varphi_0 \neq \pi/4$ should be noted.

The leading order effective strain energy is

$$\Phi^\star_{[0]} = \iint \frac{1}{2}\left((\lambda+\mu)\,\mathrm{tr}^2\,\varepsilon + 2\mu\,\varepsilon^D:\varepsilon^D\right)\,\mathrm{d}^2 X, \tag{4.16}$$

where the Lamé parameters are given by

$$(\lambda, \mu) = \left(\frac{k}{2}, \frac{k}{2}\right). \tag{4.17}$$

The effective energy is isotropic, non-degenerate and independent of $\theta$. Seeing its expression, one could (mistakenly) deduce either that (*i*) no mechanisms such as $\theta$ exist, or given $\theta$ as a kinematic field that (*ii*) any field $\theta(X)$ is a mechanism.

To see why both conclusions are inaccurate, from a macroscopic standpoint, one needs to push past the first order term $\Phi^\star_{[1]} = 0$ and include the correction at order $\ell^2$,

$$\Phi^\star_{[2]} = \iint \frac{1}{2}\left(\frac{k\,\ell^2}{4}\|\nabla\theta\|^2 + (\cdots)\,\nabla\varepsilon \otimes \nabla\varepsilon + (\cdots)\,\varepsilon \otimes \nabla^2\varepsilon\right) d^2\mathbf{X}. \quad (4.18)$$

Here too, only the term quadratic in $\nabla\theta$ is appropriately kept since any contributions from $\varepsilon$ to $\Phi^\star_{[2]}$ are dominated by contributions from $\varepsilon$ to $\Phi^\star_{[0]}$. Accordingly, the adopted effective strain energy is

$$\Phi^\star(\mathbf{u},\theta) = \iint \frac{1}{2}\left((\lambda+\mu)\,\mathrm{tr}^2\varepsilon + 2\mu\,\varepsilon^D:\varepsilon^D + \frac{k\,\ell^2}{4}\|\nabla\theta\|^2\right) d^2\mathbf{X}. \quad (4.19)$$

The effective strain energy is isotropic and kinematically enriched. The macroscopic displacement and the mechanism amplitude remain uncoupled. They can be solved for independently unless coupled extrinsically, through boundary conditions or through applied forces varying at the scale of the cell. Whether one should solve for both $\mathbf{u}$ and $\theta$ or not ultimately depends on the desired modeling level of detail, *e.g.*, is the macroscopic Cauchy stress sufficient, or is the distribution of tension in the springs also needed? For the former, $\mathbf{u}$ is enough; for the latter, $\theta$ is needed.

### 4.5. Special case of close-packed auxetic squares

We have seen that the lattice of auxetic squares, in a generic state, has an effective strain energy that is regularized by penalizing the mechanism gradient $\nabla\,\mathrm{tr}\,\varepsilon$ through a high-order elastic modulus $b(\varphi_0)$. However, this rigidity scales like $\varphi_0^2$ for $\varphi_0 \to 0$ meaning that for close-packed states, it provides little to no regularization. A similar cancellation takes place for $\varphi_0 = \pi/2$, see Figure 4.3 and Table 4.1.

Here, there are two possibilities: (*i*) regularization now needs to include even higher-order strain gradients such as $\nabla\nabla\,\mathrm{tr}\,\varepsilon$ and so on; (*ii*) no regularization by strain gradient is possible and the lattice, in the close-packed state, genuinely becomes "floppier". In the present case, it is easy enough to determine that it is (*ii*) that takes place, *i.e.*, no regularization is possible. Indeed, in the fully collapsed state, gap edges align and certain inextensibility constraints become redundant, offering the lattice more mechanisms. Thus, the effective strain energy is

$$\Phi^\star_{[0]} = \frac{1}{2} \iint 2\mu\,\varepsilon^D:\varepsilon^D \, d^2\mathbf{X}, \quad (4.20)$$

and all conformal transformations are now *mechanisms* rather than *soft modes*. This is similar to how the square lattice has, and maintains at all orders, a degenerate effective energy reflecting the fact that the lattice admits an abundance of arbitrarily graded pure shear mechanisms. That being said, the mechanisms of the square lattice can be pursued into nonlinear mechanisms whereas the mechanisms of the close-packed auxetic squares become soft modes as soon as the nonlinearity kicks in accordance with the generic case treated in (4.3).

### 4.6. Kagome lattice (generic case)

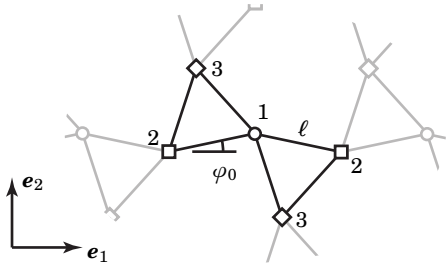

**Figure 4.4.** Reference configuration of the kagome lattice, parameterized by the amplitude $\varphi_0$ of the mechanism.

The kagome lattice in a generic, pre-twisted, state is depicted on Figure 4.4. As with the auxetic squares, the reference configuration is parameterized by a pre-twist angle $\varphi_0$, which we take in the range $0 \leqslant \varphi_0 \leqslant \pi/2$ as the undeformed configuration is globally invariant by the transformation $\varphi_0 \leftarrow \pi - \varphi_0$. We also exclude the particular value $\varphi_0 = \pi/2$, as the lattice then shrinks to a triangle, and there is no separation of scales.

The unit cell problem is regular for all values of $\varphi_0 \in [0, \pi/2)$, except for the particular value $\varphi_0 = 0$ corresponding to full extension, which is studied separately in subsection 4.7. For all other values $0 < \varphi_0 < \pi/2$, the effective strain energy, to leading order, is

$$\Phi^\star_{[0]} = \frac{1}{2} \iint (2\mu)\,\varepsilon^D:\varepsilon^D \, d^2\mathbf{X}, \quad (4.21)$$

where the shear modulus is $\mu = \frac{\sqrt{3}\,k}{8}$. The underlying microscopic shift is of a similar form as for auxetic squares, see (4.10)–(4.11),

$$\mathbf{y}_0 = \underbrace{-\frac{1}{2\tan\varphi_0}\,\mathrm{tr}\,\varepsilon \times \mathbf{y}^\theta(\varphi_0)}_{\text{twist angle}} + (\cdots)\,\varepsilon^D, \quad (4.22)$$

and the normalized twisting mode (*i.e.*, with alternated triangle rotations having angle $\pm 1$) is now

$$\boldsymbol{y}^\theta(\varphi_0) = \frac{1}{\sqrt{3}\cos\varphi_0}\left(\begin{array}{ccc} \frac{\boldsymbol{e}_1+\sqrt{3}\boldsymbol{e}_2}{2} & \frac{\boldsymbol{e}_1-\sqrt{3}\boldsymbol{e}_2}{2} & -\boldsymbol{e}_1 \end{array}\right). \tag{4.23}$$

The twisting corresponds to an increment of the initial angle $\varphi_0$ in Figure 4.4, and gives rise macroscopically to an isotropic dilation.

We return to the energy (4.21). It is degenerate and vanishes over all conformal transformations. By contrast to the case of auxetic squares, this account of the mechanisms of the kagome lattice is in fact quite faithful. Indeed, the kagome lattice is very "floppy" and accommodates not only uniform twisting, but also its gradient, as mechanisms.

Suppose, however, that we did not have such a priori knowledge of the "floppiness" of the kagome lattice. By the general approach proposed in the decision tree in Figure 2.1, one would then try to regularize $\Phi^\star_{[0]}$ by including strain gradients, similar to what was done for auxetic squares. Doing so, one obtains the correction at order $\ell^1$ as

$$\Phi^\star_{[1]} = \iint \mu\,\ell\,(a(\varphi_0)\,\nabla(\mathrm{tr}\,\boldsymbol{\varepsilon})\cdot\boldsymbol{\mathcal{J}}:\boldsymbol{\varepsilon}^D + (\cdots)\,(\boldsymbol{\varepsilon}^D\otimes\nabla\boldsymbol{\varepsilon}^D))\,\mathrm{d}^2\boldsymbol{X} \tag{4.24}$$

where the coefficient $a(\varphi_0)$ is given by

$$a(\varphi_0) = \frac{\cos(2\varphi_0)}{2\sqrt{3}\sin\varphi_0}, \tag{4.25}$$

and $\boldsymbol{\mathcal{J}}$ is a fixed tensor of rank 3 relevant to the $D_6$ symmetry of the kagome lattice.

$$\boldsymbol{\mathcal{J}} = \boldsymbol{e}_1\otimes(-\boldsymbol{e}_1\otimes\boldsymbol{e}_1+\boldsymbol{e}_2\otimes\boldsymbol{e}_2)+\boldsymbol{e}_2\otimes(\boldsymbol{e}_1\otimes\boldsymbol{e}_2+\boldsymbol{e}_2\otimes\boldsymbol{e}_1), \tag{4.26}$$

The correction at order $\ell^2$ is

$$\Phi^\star_{[2]} = \frac{1}{2}\iint \mu\,\ell^2\left(\begin{array}{c} a^2(\varphi_0)\,\|\nabla(\mathrm{tr}\,\boldsymbol{\varepsilon})\|^2 + (\,\ldots\,)\,\nabla(\mathrm{tr}\,\boldsymbol{\varepsilon})\otimes\nabla\boldsymbol{\varepsilon}^D + (\,\ldots\,)\,\nabla\boldsymbol{\varepsilon}^D\otimes\nabla\boldsymbol{\varepsilon}^D \\ +\,(\,\ldots\,)\mathrm{tr}\,\boldsymbol{\varepsilon}\,\nabla^2\boldsymbol{\varepsilon}^D + (\,\ldots\,)\,\boldsymbol{\varepsilon}^D\otimes\nabla^2\boldsymbol{\varepsilon} \end{array}\right)\mathrm{d}^2\boldsymbol{X}. \tag{4.27}$$

Focus now on soft modes that are conformal, or nearly conformal, *i.e.*, on the regime where $\boldsymbol{\varepsilon}^D \sim \ell\,\nabla(\mathrm{tr}\,\boldsymbol{\varepsilon})$. Upon noting that $\boldsymbol{\varepsilon}^D:\boldsymbol{\varepsilon}^D = \|\boldsymbol{\mathcal{J}}:\boldsymbol{\varepsilon}^D\|^2/2$, the effective strain energy reduces to

$$\begin{aligned} \Phi^\star &= \iint \mu\left(\boldsymbol{\varepsilon}^D:\boldsymbol{\varepsilon}^D + \ell a\,\nabla(\mathrm{tr}\,\boldsymbol{\varepsilon})\cdot\boldsymbol{\mathcal{J}}:\boldsymbol{\varepsilon}^D + \frac{1}{2}\ell^2 a^2\,\|\nabla(\mathrm{tr}\,\boldsymbol{\varepsilon})\|^2\right)\mathrm{d}^2\boldsymbol{X}\\ &= \iint \frac{1}{2}\mu\left\|\boldsymbol{\mathcal{J}}:\boldsymbol{\varepsilon}^D + a(\varphi_0)\,\ell\,\nabla(\mathrm{tr}\,\boldsymbol{\varepsilon})\right\|^2\mathrm{d}^2\boldsymbol{X}. \end{aligned} \tag{4.28}$$

Ultimately, then, penalizing the twisting gradient $\nabla(\mathrm{tr}\,\boldsymbol{\varepsilon})$ does not suppress conformal maps but simply *corrects* them: regularization by strain gradients does not take place. This is consistent with the aforementioned floppiness.

Incidentally, the "post-conformal" corrected field equation of the mechanisms is

$$\boldsymbol{\mathcal{J}}:\boldsymbol{\varepsilon}^D + a(\varphi_0)\,\ell\,\nabla(\mathrm{tr}\,\boldsymbol{\varepsilon}) = \boldsymbol{0}. \tag{4.29}$$

**Remark 4.1.** The tensor $\boldsymbol{\mathcal{J}}$ introduced in (4.26) is zero on dilations, $\boldsymbol{\mathcal{J}}:\boldsymbol{I}_2 = \boldsymbol{0}$. As a result, we will systematically replace $\boldsymbol{\mathcal{J}}:\boldsymbol{\varepsilon}$ with $\boldsymbol{\mathcal{J}}:\boldsymbol{\varepsilon}^D$. In effect, $\boldsymbol{\mathcal{J}}$ casts the deviatoric part of the strain into a vector $\boldsymbol{\mathcal{J}}:\boldsymbol{\varepsilon}^D \in \mathbb{R}^2$.

### 4.7. Special case of a fully extended kagome lattice

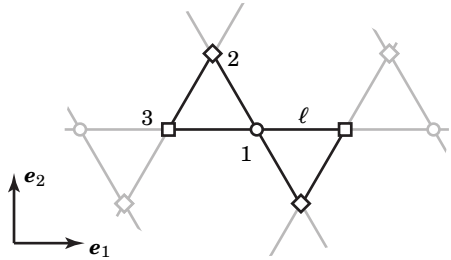

**Figure 4.5.** Reference configuration of the fully extended kagome lattice ($\varphi_0 = 0$).

When the kagome lattice is fully extended, $\varphi_0 \to 0$, the coefficient $\frac{1}{2\tan\varphi_0}$ relating $\mathrm{tr}\,\boldsymbol{\varepsilon}$ to the amplitude of the twisting mode diverges, see (4.22). In close analogy with the fully extended auxetic squares, this divergence arises because the microscopic mode produces a macroscopic dilation with a smaller and smaller 'efficiency'. As a result, the twisting angle diverges as well unless $\mathrm{tr}\,\boldsymbol{\varepsilon}$ approached zero at the same time. But then $\frac{1}{2\tan\varphi_0}\mathrm{tr}\,\boldsymbol{\varepsilon} \sim \frac{\mathrm{tr}\,\boldsymbol{\varepsilon}}{2\varphi_0}$ becomes indeterminate. This amplitude takes a life of its own when $\varphi_0 = 0$ and becomes a new kinematic field $\theta$.

The additional kinematic field is added automatically by the homogenization "recipe", which detects that the cell problem is singular for $\varphi_0 = 0$, with a rank deficiency of 1. The degrees of freedom $\boldsymbol{l}$ and the microscopic shift are obtained in the same forms (4.14) and (4.15), respectively, as for the fully extended auxetic squares,

$$\boldsymbol{y}_0 = \boldsymbol{y}_0(\boldsymbol{l}(\boldsymbol{X})) = \theta(\boldsymbol{X})\underbrace{\frac{1}{\sqrt{3}}\left(\begin{array}{ccc} \frac{\boldsymbol{e}_1+\sqrt{3}\boldsymbol{e}_2}{2} & \frac{\boldsymbol{e}_1-\sqrt{3}\boldsymbol{e}_2}{2} & -\boldsymbol{e}_1 \end{array}\right)}_{\boldsymbol{y}^\theta(0)} + (\cdots)\,\boldsymbol{\varepsilon}^D(\boldsymbol{X}). \tag{4.30}$$

The leading-order effective strain energy is

$$\Phi^\star_{[0]} = \iint \frac{1}{2}\left((\lambda+\mu)\operatorname{tr}^2\varepsilon + 2\mu\varepsilon^D:\varepsilon^D\right) d^2X \tag{4.31}$$

with Lamé parameters $\lambda = \mu = \frac{\sqrt{3}k}{8}$. The energy $\Phi^\star_{[0]}$ is non-degenerate and is the first example, thus far, of an effective strain energy that *over-estimates* the rigidity of the lattice by omitting mechanisms. Specifically, $\Phi^\star_{[0]}$ over-estimates the shear rigidity. Indeed, not only does $\Phi^\star_{[0]}$ eliminate mechanism $\theta$ but also eliminates any chance for $\theta$ and its gradient, to relax a macroscopic strain $\varepsilon$. It is actually possible to relax, or conversely to produce, a *deviatoric* macroscopic strain $\varepsilon$ in the lattice using a twist gradient $\nabla\theta$ of order $\varepsilon/\ell$. Here is how.

Consider the first- and second-order corrections

$$\Phi^\star_{[1]} = \mu\left(-\frac{1}{\sqrt{3}}\iint (\mathscr{J}:\varepsilon^D)\cdot(\ell\,\nabla\theta)\, d^2X\right) \tag{4.32}$$

and

$$\Phi^\star_{[2]} = \iint \frac{1}{2}\mu\left(\frac{1}{3}\|\ell\,\nabla\theta\|^2 + (\ldots)\,(\nabla\varepsilon\otimes\nabla\varepsilon) + (\ldots)\,(\varepsilon\otimes\nabla^2\varepsilon)\right) d^2X. \tag{4.33}$$

Consistent with the proposed scaling, namely $\varepsilon\sim\ell\,\nabla\theta$, we retain in $\Phi^\star_{[2]}$ the quadratic terms in $\nabla\theta$ only, leading to a total effective strain energy

$$\Phi^\star = \iint \frac{1}{2}\left((\lambda+\mu)\operatorname{tr}^2\varepsilon + \mu\left\|\mathscr{J}:\varepsilon^D - \frac{\ell\,\nabla\theta}{\sqrt{3}}\right\|^2\right) d^2X. \tag{4.34}$$

Thus, mechanisms exist and are solutions to

$$\operatorname{tr}\varepsilon = 0, \quad \mathscr{J}:\varepsilon^D - \frac{\ell\,\nabla\theta}{\sqrt{3}} = 0. \tag{4.35}$$

In particular, for any *uniform* deviatoric strain $\varepsilon = \varepsilon^D$, there exists a *linear* twist $\theta$ such that $\Phi^\star = 0$. By contrast, in the fully extended lattice of auxetic squares, neither $\theta$ nor its gradient were able to relax a deviatoric strain $\varepsilon^D$ which is yet another manifestation of the rigidity of auxetic squares relative to the kagome lattice.

## 5. CROSS-OVER REGIMES

The generic effective models do not match up, in general, with those relevant to a specific initial configuration. For the auxetic squares, for instance, we obtained in the generic case (4.9) and for the fully extended configuration (4.19), respectively,

$$\Phi^\star = \begin{cases} \frac{k}{2}\iint\left(\quad\quad\quad\varepsilon^D:\varepsilon^D + b(\varphi_0)\|\ell\,\nabla\operatorname{tr}\varepsilon\|^2\quad\right) d^2X & \text{if } \varphi_0 \neq \pi/4 \\ \frac{k}{2}\iint\left(\operatorname{tr}^2\varepsilon + \varepsilon^D:\varepsilon^D + \tfrac{1}{4}\|\ell\,\nabla\theta\|^2\quad\right) d^2X & \text{if } \varphi_0 = \pi/4 \end{cases} \tag{5.1}$$

By (4.8), the higher-order modulus $b(\varphi_0)$ diverges for $\varphi_0\to\pi/4$ and thus does not match with the value $1/4$ obtained for $\varphi_0 = \pi/4$. Even more significantly, the $\operatorname{tr}^2\varepsilon$ term is entirely absent from the generic model.

In this section, we propose an approach to identify an effective model covering both the generic case $\varphi_0 \neq \pi/4$ and particular case $\varphi_0 = \pi/4$, using the auxetic squares as an illustration. The effective model thus obtained captures a new *cross-over* regime, when the deviation of $\varphi_0$ from the critical value $\pi/4$ is of the same order of magnitude as the scale separation parameter $\gamma = \ell/L$, i.e., $|\varphi_0 - \pi/4| = \mathcal{O}(\gamma) \ll 1$, see Equation (5.8) below. This cross-over regime has been overlooked in the above analyses, as we implicitly assumed that $\varphi_0$ was independent of $\gamma$. It can be obtained by a simple extension of the homogenization method, which we proceed to present.

What makes the two versions of the effective models in (5.1) difficult to reconcile is that the enrichment variable $\theta$ is only added *automatically* by the homogenization "recipe" when the cell problem is singular, for $\varphi_0 = \pi/4$. The library still allows enrichment variables to be introduced *manually*, and this is the strategy we use to derive the effective model applying to the cross-over regime. Along with the definition of the lattice passed to the library shoal, we include an option kinematicEnrichment that instructs the library to append a variable $\theta(X)$ to the macroscopic degrees of freedom—as was done earlier in (4.14), but now for *any* value of $\varphi_0$—, and to connect it with the microscopic shift by

$$\forall X, \quad \underbrace{\xi_I(X)\cdot e_1 + \xi_{II}(X)\cdot e_2 - \xi_{III}(X)\cdot e_1 - \xi_{IV}(X)\cdot e_2}_{y(X)\cdot(+e_1\ +e_2\ -e_1\ -e_2)} = \frac{4}{\cos\varphi_0 + \sin\varphi_0}\theta(X). \tag{5.2}$$

This ensures that $\theta(X)$ is the amplitude of the twisting mode, see (4.10), (4.11) and (2.7). The kinematic constraint (5.2) is dealt with as follows: the residual $(E_{[l,y]})_{n_\varphi+3} = y(X)\cdot(+e_1\ +e_2\ -e_1\ -e_2) - 4\theta(X)/(\cos\varphi_0 + \sin\varphi_0)$ of (5.2) is added to the list of the microscopic degrees of freedom $E$, its dependence on the macroscopic and microscopic degrees of freedom is worked out as $E_{[l,y]}(X) = E_l\cdot l(X) + \gamma E'_l:\nabla l(X) + \cdots + E_y\cdot y(X) + \gamma E'_y:\nabla y(X) + \cdots$, see (2.10), and the matrix $\mathcal{Q}$ is extended so that Equation (5.2) is included in the set of kinematic constraints (2.12).

Having set up enrichment for the micro-twist variable $\theta$, we proceed to run automatic homogenization. The cell problem is found to be always regular, even for $\varphi_0 = \pi/4$. This is not surprising as we have precisely used the enrichment variable $\theta$ that has been used to fix the singular case $\varphi_0 = \pi/4$. The effective energy at leading order is

$$\Phi^\star_{[0]}[\varepsilon,\theta] = \iint \frac{k}{2}\left(\frac{(\operatorname{tr}\varepsilon + 2\tan(\varphi_0 - \pi/4)\,\theta)^2}{\cos^2(\varphi_0 - \pi/4)} + \varepsilon^D:\varepsilon^D\right) d^2X \tag{5.3}$$

The model is free of any singularity, as all the coefficients are continuous (except for the fully collapsed configuration $\varphi_0 = {}^{3\pi}/_4$ which we have excluded from the onset). The solution $\boldsymbol{\varepsilon} = -\boldsymbol{I}_2 \tan(\varphi_0 - {}^\pi/_4)\,\theta$ has a zero energy: this is the mechanism when $\varphi_0 \neq {}^\pi/_4$, and it degenerates into a purely microscopic mode ($\boldsymbol{\varepsilon} = \boldsymbol{0}$, $\theta \neq 0$) when $\varphi_0 = {}^\pi/_4$.

The presence of this zero mode in the leading-order energy suggests that homogenization should be pushed to higher order. Due to centro-symmetry of the lattice, the order-$\ell$ correction vanishes, $\Phi^\star_{[1]} = 0$. By including the order-$\ell^2$ correction, we obtain an effective model in the form

$$\Phi^\star = \iint \frac{k}{2}\left(\frac{(\operatorname{tr}\boldsymbol{\varepsilon} + 2\tan(\varphi_0 - {}^\pi/_4)\,\theta)^2}{\cos^2(\varphi_0 - {}^\pi/_4)} + \boldsymbol{\varepsilon}^D : \boldsymbol{\varepsilon}^D + \ell^2 c(\varphi_0)\|\nabla\theta\|^2\right) \mathrm{d}^2\boldsymbol{X}, \tag{5.4}$$

where the coefficient $c(\varphi_0)$ is given by

$$c(\varphi_0) = \frac{1}{4}\frac{\cos^2(2(\varphi_0 - {}^\pi/_4))}{\cos^2(\varphi_0 - {}^\pi/_4)}. \tag{5.5}$$

As we have done many times earlier, we have dropped in (5.4) gradient terms proportional to $\nabla \boldsymbol{\varepsilon}^D$ and $\nabla(\operatorname{tr}\boldsymbol{\varepsilon} + 2\tan(\varphi_0 - {}^\pi/_4)\,\theta)$, corresponding to the quantities that are already penalized at leading order. We refer the reader to the companion Mathematica notebook [4] for a detailed derivation of (5.4).

The effective model (5.4) can be used indifferently with $\varphi_0 = {}^\pi/_4$ or $\varphi_0 \neq {}^\pi/_4$, and the previous behaviors are recovered as particular cases:

- For $\varphi_0 \neq {}^\pi/_4$, one can optimize out $\theta$ in the leading-order term, which yields again the micro-twist amplitude $\theta = -\operatorname{tr}\boldsymbol{\varepsilon}/(2\tan(\varphi_0 - {}^\pi/_4))$ of the mechanism. Inserting this into the gradient term and using the identity

$$c(\varphi_0)\left(-\frac{1}{2\tan(\varphi_0 - {}^\pi/_4)}\right)^2 = b(\varphi_0), \tag{5.6}$$

which is proven in the companion notebook [4], one recovers the generic effective model $(5.1)_1$ from Section 4.3.

- For $\varphi_0 = {}^\pi/_4$, the $\theta$ variable disappears from the leading-order term and a penalization of the form $(\operatorname{tr}\boldsymbol{\varepsilon})^2$ takes place. Disregarding the corresponding gradient term $\nabla(\operatorname{tr}\boldsymbol{\varepsilon})$ and using $c({}^\pi/_4) = {}^1/_4$, see (5.5), one recovers the effective model $(5.1)_2$ for the fully extended case from Section 4.4.

By limiting attention to values of $\varphi_0$ that are close to ${}^\pi/_4$ and by simplifying the coefficients accordingly, $\tan(\varphi_0 - {}^\pi/_4) \approx \varphi_0 - {}^\pi/_4$ and $c(\varphi_0) \approx {}^1/_4$, one can obtain slightly simpler version of the effective model (5.4) for nearly-extended auxetic squares,

$$\Phi^\star = \iint \frac{k}{2}\left((\operatorname{tr}\boldsymbol{\varepsilon} + 2(\varphi_0 - {}^\pi/_4)\,\theta)^2 + \boldsymbol{\varepsilon}^D : \boldsymbol{\varepsilon}^D + \frac{\ell^2}{4}\|\nabla\theta\|^2\right) \mathrm{d}^2\boldsymbol{X}. \tag{5.7}$$

With $\gamma = \ell/L \ll 1$ as the scale separation parameter, this effective energy is associated with the scaling assumptions

$$|\varphi_0 - {}^\pi/_4| = \mathcal{O}(\gamma), \qquad \boldsymbol{\varepsilon} \sim \theta\gamma, \tag{5.8}$$

and yields again the two limit cases in (5.1) as particular cases—with the coefficient $b(\varphi_0)$ replaced by its asymptotically equivalent $b(\varphi_0) \sim {}^1/_{(16(\varphi_0 - {}^\pi/_4)^2)}$, however.

## 6. Conclusion and perspectives

The classification procedure proposed in the present paper provides a unified view of the different classes of effective models that can be obtained by homogenizing soft lattices: standard elasticity (as for the triangular lattice), standard elasticity with macroscopic constraint (as for the honeycomb with inextensible beams), strain gradient elasticity (as for the auxetic squares in the generic case), enriched continuum (as for the fully extended auxetic squares and kagome lattices). The variety of models has been shown to emerge from a common pattern of analysis based on asymptotic second-order homogenization. We provide an implementation of the symbolic homogenization engine in the open-source library shoal. It is an extension of the version published in an earlier work, that can handle periodic mechanisms. We have demonstrated the procedure on a range of examples involving networks of pin-jointed springs in 2D. The library is quite flexible and can handle equally 3D lattices, networks of beams and/or to elastic joints. Our approach automatically identifies periodic mechanisms whose spatial period matches the period of the lattice. Periodic mechanisms spreading over multiple periodic unit cells can be handled, by running the procedure on an 'extended' cell which needs to be identified ahead of time. As it is based on homogenization, our procedure does not account for *localized* mechanisms whose amplitude varies from on cell to the next.

In the last section of the paper, we have show that it is possible to use manual enrichment to derive enriched models capturing the cross-over regime, *i.e.*, the progressive transition from a regular effective behavior to a critical effective behavior at a particular value of the lattice parameters when, *e.g.*, a microscopic mechanism appears. In future work, it would be interesting to explore how this manual enrichment procedure could be used to introduce a relaxation parameter capturing the effect of the gradient of the mechanism amplitude, without the need of a proper strain gradient model. This could be beneficial from a numerical viewpoint.

In real elastic microstructures, such as kirigami materials that are more faithfully represented as a network of 2D or 3D elastic bodies than 1D springs or beams, thin elastic junctions replace idealized joints, and the (zero-energy) mechanisms are turned into (low-energy) soft modes. In these systems, gradient elasticity and enriched models are very efficient tools to capture long-range modes of interactions in an effective way. It would be of great practical interest to extend the proposed procedure of analysis to this class of structures.